\documentclass[a4paper,10pt]{article}
\usepackage[utf8]{inputenc}
\usepackage{amssymb,amsmath}
\usepackage{authblk}
\usepackage[left=0.5in, right=0.5in, top=1in, bottom=1in]{geometry}
\usepackage{hyperref}
\usepackage{graphicx}

\title{Energy fluctuation of ideal Fermi gas trapped under generic power 
law potential $U=\sum_{i=1} ^d c_i |\frac{x_i}{a_i}|^{n_i}$
in d dimension}

\author{Mir Mehedi Faruk$^{a,b}$, Md. Muktadir Rahman$^{b}$, 
Dwaipayan Debnath $^{b}$, Sakhawat Himel$^{b}$\\
Theoretical Physics, Blackett Laboratory, Imperial College, London SW7 2AZ, United Kingdom$^a$\\
Department of Theoretical Physics, University of Dhaka, Dhaka-1000$^b$\\

\href{mailto:me@somewhere.com}{Email: muturza3.1416@gmail.com, mir.faruk15@imperial.ac.uk, mir.mehedi.faruk@cern.ch} 
 }

\begin{document}
\maketitle
 
 \begin{abstract}
 Energy fluctuation of ideal Fermi gas trapped under generic power 
law potential $U=\sum_{i=1} ^d c_i |\frac{x_i}{a_i}|^{n_i}$ have been calculated
in arbitrary dimension. Energy fluctuation is scrutinized further 
in the degenerate limit $\mu>> K_BT$ with the help of Sommerfeld expansion. The dependence of energy fluctuation on dimensionality
and power law potential is studied in detail. Most importantly our 
general result can exactly reproduce the recently published result regarding free and harmonically trapped ideal
Fermi gas
in d=3\cite{s}.
\end{abstract}
\Large
\section{Introduction}
A lot of theoretical studies\cite{bose,einstein,pathria,ziff,howard,may} have  been done on the subject of
ideal free quantum gases even before the experimental observation
of Bose-Einstein condensation (BEC) and Fermi degeneracy. 
But, this subject drew more attention after it had been possible to detect  BEC\cite{Bradley,anderson,davis}
and Fermi degeneracy\cite{DeMarco} experimentally
in trapped quantum gases.
 Since then, an increasing attraction is noticed in the subject of
 trapped quantum gases.
Although a lot of theoretical 
studies\cite{s,sala,yan,yan1,yan2,yan3,marek1,turza,t2}
are done on quantum gases trapped under generic power law potential, $U=\sum_{i=1} ^d c_i |\frac{x_i}{a_i}|^{n_i}$,
none of these contained
detailed discussion on energy fluctuation $\Delta \epsilon^2$,
until the recent paper of Biswas et. al.\cite{s} 
where they discussed the energy fluctuation for free and harmonically 
trapped quantum gases in three dimensional space. 
In their paper they have also
conjectured a relation between 
discontinuity of $C_V$ and energy fluctuation $\Delta \epsilon^2$, 
suggesting 
the appearance of a hump in $\frac{\Delta \epsilon^2}{ kT^2}$  over its classical limit 
does
indicate a discontinuity of $C_V$.
It was also reported in their study that, there is no hump
in $\frac{\Delta \epsilon^2}{ kT^2}$, in the case of free and harmonically trapped ideal
Fermi gases. In their recent paper, Mehedi et. al. has proved the conjecture and investigated energy fluctuation\cite{mms}
in details for ideal Bose gas trapped under generic power law potential
in arbitrary dimension.
Point to note that, Biswas et. al. calculated the $\Delta \epsilon^2$ for three dimensional
free and harmonically trapped quantum gases, 
but  $\Delta \epsilon^2$ is still not examined in arbitrary dimension while Fermi gas is trapped under generic
power law potential.\\\\
In this paper we investigate the $\Delta \epsilon^2$ of ideal Fermi gases trapped under generic
power law potential $U=\sum_{i=1} ^d c_i |\frac{x_i}{a_i}|^{n_i}$
in $d$ dimension. At first we determine the density of states, which enables us to calculate the 
energy fluctuation using the
Fermi distribution function. Later, we scrutinize the energy fluctuation in the quantum degenerate limit using the Sommerfeld
expansion. The dependence of energy fluctuation on
dimensionality and power law exponents are visited in detail.
Interestingly our more general final result of energy fluctuation can
exactly reproduce the same result
in three dimension, for
free and harmonically trapped Fermi gases
reported in Biswas et. al. \cite{s}
by choosing $d=3$, $n=\infty$ (free Fermi gas in three dimension)
and $d=3$, $n=2$ (harmonically trapped
Fermi gas in three dimension).

\section{Energy fluctuation of trapped Fermi gas} 
Considering an ideal  gas trapped under a generic power law potential
in $d$ dimensional space with  a single particle hamiltonian\cite{yan},
\begin{eqnarray}
 \epsilon (p,x_i)= bp^l + \sum_{i=1} ^d c_i |\frac{x_i}{a_i}|^{n_i}
\end{eqnarray}
where, $p$ is the momentum 
and $x_i$ is the  $i$ th component of coordinate of a particle
and 
$b,$ $l,$ $a_i$, $c_i$, $n_i$  
are defined as all positive constants.
Here, $c_i$, $a_i$, $n_i$ determine the depth 
and confinement power of
the potential and $l$ being the kinematic parameter, where $x_i<a_i$.
As $|\frac{x_i}{a_i}|<1$,
the potential term goes to zero
as all $n_i\longrightarrow \infty$.
We can construct our usual non-relativistic Hamiltonian
with $l=2$ and $b=\frac{1}{2m}$.
The density of states for such system is \cite{yan2,t2},
\begin{eqnarray}
 \rho(\epsilon)=C(m,V_d ')\epsilon^{\chi-1}
\end{eqnarray}
where, $C(m,V_d')$ is a constant depending on effective volume $V_d'$\cite{yan1,yan3}\footnote{to read a detail discussion
on effective volume, see \cite{yan,yan1,yan2,yan3}} 
and $\chi=\frac{d}{l}+\sum_i ^d \frac{1}{n_i}$.
Now the Fermi distribution function, is given by
\begin{eqnarray}
\bar{n}_i &=&\frac{1}{z^{-1}e^{\beta\epsilon_i}+1}
\end{eqnarray}\\
where, $z$ is fugacity.
So, the energy fluctuation of trapped fermi gas,
\begin{eqnarray}
 \Delta \epsilon^2&=&\bar{\epsilon^2}-\bar{\epsilon}^2=\sum_i \bar{n}_i \epsilon_i ^2 - (\sum_i \bar{n}_i \epsilon_i)^2
 =\int d\epsilon \rho(\epsilon) \epsilon^2 n(\epsilon)  - (\int d\epsilon \rho(\epsilon)  \epsilon n(\epsilon) )^2\nonumber \\
&=& (kT)^2[\chi(\chi+1)\frac{f_{\chi+2}(\sigma)}{f_{\chi}(\sigma)}-\chi^2\frac{f_{\chi+1} ^2(\sigma)}{f_{\chi} ^2(\sigma)}]
 \end{eqnarray}
where, $f_p(z)$ is the Fermi function defined as,
\begin{equation}
  f_p(z)=\int_0 ^\infty dx\frac{x^{p-1}}{z^{-1}e^x+1}=\sum_{j=1} ^\infty (-1)^{j-1}\frac{z^j}{j^p}
\end{equation}

 \section{Energy fluctuation of trapped Fermi gas in the degenerate limit} 
   At low temperature, we can approximate the Fermi
function and write it as quickly convergent Sommerfeld series\cite{pathria}
\begin{eqnarray}
 f_p(z)=\frac{(\ln z) ^p}{\Gamma(p+1)}[1+p(p-1)\frac{\pi^2}{6}\frac{1}{(\ln z)^2}+p(p-1)(p-2)(p-3)\frac{7 \pi^4}{360}\frac{1}{(\ln z)^4}+...]
\end{eqnarray}
From Ref.\cite{t2} we can write
  the chemical potential (fugacity)  as below,
\begin{eqnarray}
\mu=kT\ln z=E_F[1-(\chi-1)\frac{\pi^2}{6}(\frac{kT}{E_F})^2] 
\end{eqnarray}
  The expression of Fermi energy for Fermi gas trapped under generic power law potential can be found in Ref. \cite{t2}.
 So, using the Sommerfeld approximation we can re-write the energy fluctuation from eq. (4),
\begin{equation}
 \Delta \epsilon^2=\frac{\chi}{(\chi+2)(\chi+1)}(kT\ln z)^2+\frac{\pi^2}{3}\frac{\chi(2\chi+1)}{\chi+2}(kT)^2
 -\frac{2\pi^2}{3}\frac{\chi^2}{(\chi+1)^2}(kT)^2
\end{equation}
Again using of Eq. (7), the energy fluctuation becomes
\begin{equation}
 \frac{\Delta \epsilon^2}{E_F ^2}=
 \frac{\chi}{(1+\chi)^2(2+\chi)}+ \frac{1}{3}\frac{1}{(1+\chi)^2}\pi^2\tau^2+\frac{1+34\chi+40\chi^2-8\chi^3-13\chi^4
 +2\chi^6}{36(1+\chi)^2(2+\chi)}\pi^4\tau^4
\end{equation}
where, $\tau=\frac{T}{T_F}$.
At $T=0$, the energy fluctuation becomes,
\begin{equation}
 \Delta \epsilon_0 ^2=\frac{ \chi}{(\chi+2)(\chi+1)}E_F^2
 \end{equation}
In the case of ideal free Fermi gas in three dimensional space, $\chi=3/2$. So, from Eq. (4)
we see $\Delta \epsilon^2$ becomes,
\begin{eqnarray}
\frac{\Delta \epsilon_0 ^2}{E_F ^2}=\frac{12}{175}+\frac{\pi^2}{5}\tau^2+\frac{329\pi^4}{2400}\tau^4
\end{eqnarray}
And when the ideal Fermi gas is trapped under a harmonic potential in three dimension, ($\chi=3$)
\begin{eqnarray}
\frac{\Delta \epsilon_0 ^2}{E_F ^2}=\frac{3}{80}+\frac{\pi^2}{4}\tau^2+\frac{163\pi^4}{240}\tau^4
\end{eqnarray}
Equation (11) and (12) coincides exactly with Biswas et. al. \cite{s}

\section{Results and Discussion}
In this section we summarize the interesting findings relating
energy fluctuation of ideal Fermi gas trapped under generic power law potential.\\\\
\begin{figure}[h!]
\centering
\includegraphics[ height=8cm, width=10cm]{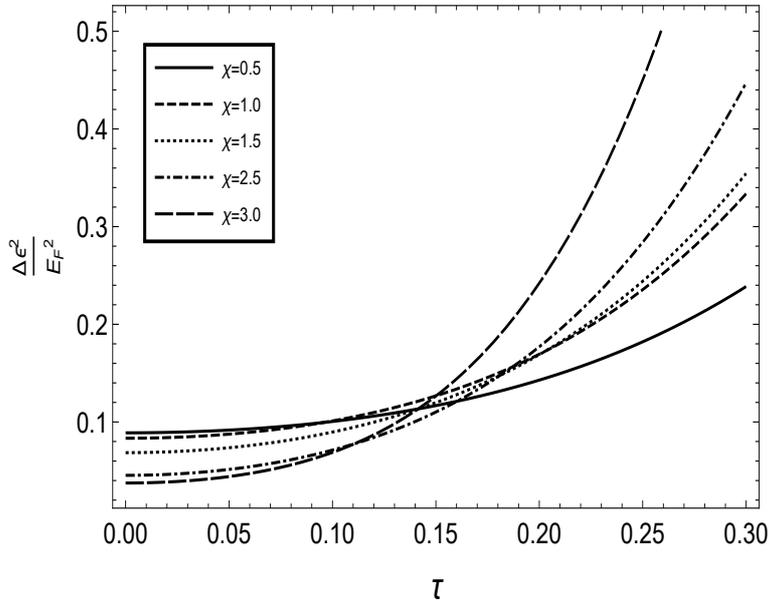}  
\caption{Energy fluctuation ideal trapped Fermi gas
as a function of $\tau=\frac{T}{T_F}$, with different power law potentials. }
  \label{fig:boat1}
\end{figure}\\\\
It is seen in the studies that\cite{ziff}, 
all the thermodynamic quantities of free Fermi (Bose)
gases can be presented in terms
of Fermi function (Bose function) depending on dimensionality $d$.
Now, the thermodynamic 
quantities
of trapped Fermi (Bose) gases 
can still be written in terms of Fermi function (Bose function), using the concept 
of effective volume and effective thermal wavelength\cite{yan,t2}. But in this case  
the Fermi (Bose) functions depend on $\chi=\frac{d}{l}+\sum_i ^d \frac{1}{n_i}$.
So, at first we explore the dependence of energy fluctuation
on $\tau=\frac{T}{T_F}$, with varying $\chi$ (figure 1).
Here, all $n_i\longrightarrow \infty$ correspond to free system\cite{yan}.
It is very enthralling to point out that,
the energy fluctuation is non-zero at $T=0K$ for any value of $\chi$,
 unlike the Bose gas for which the 
 \begin{figure}[h!]
\centering
\includegraphics[ height=8cm, width=8cm]{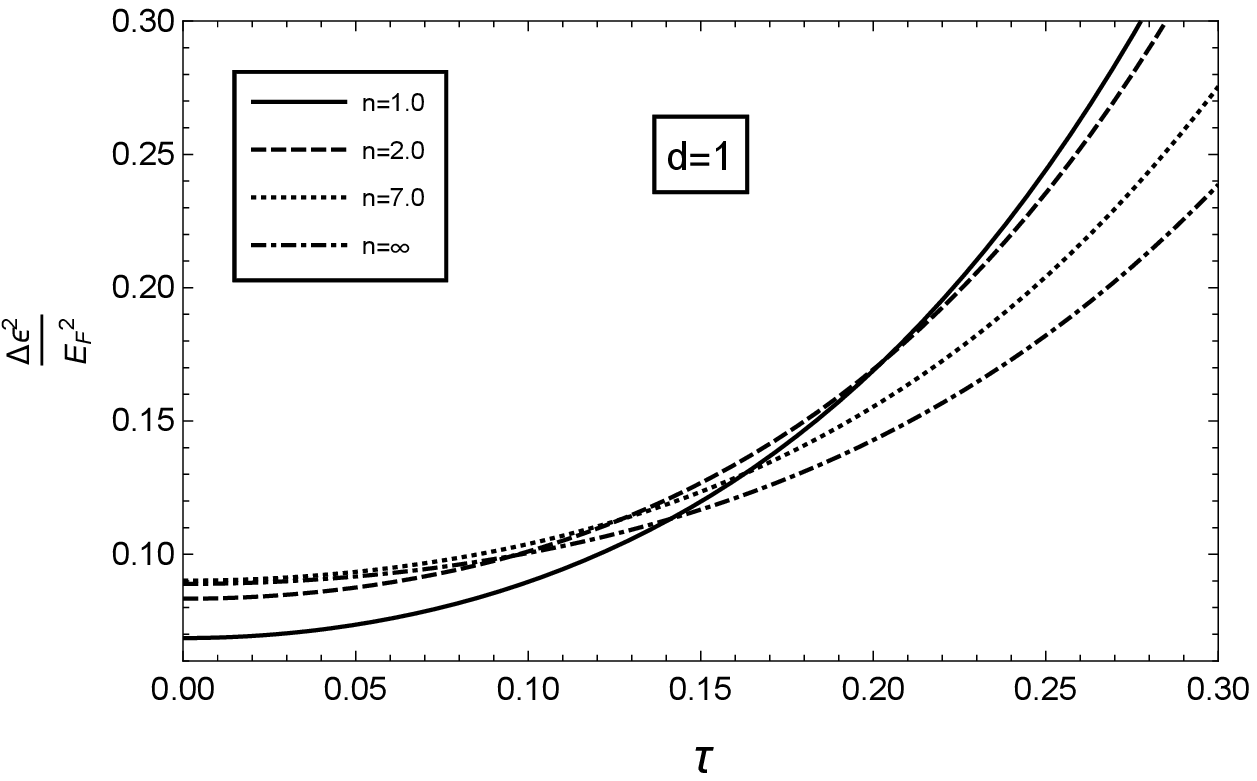}  
\includegraphics[ height=8cm, width=8cm]{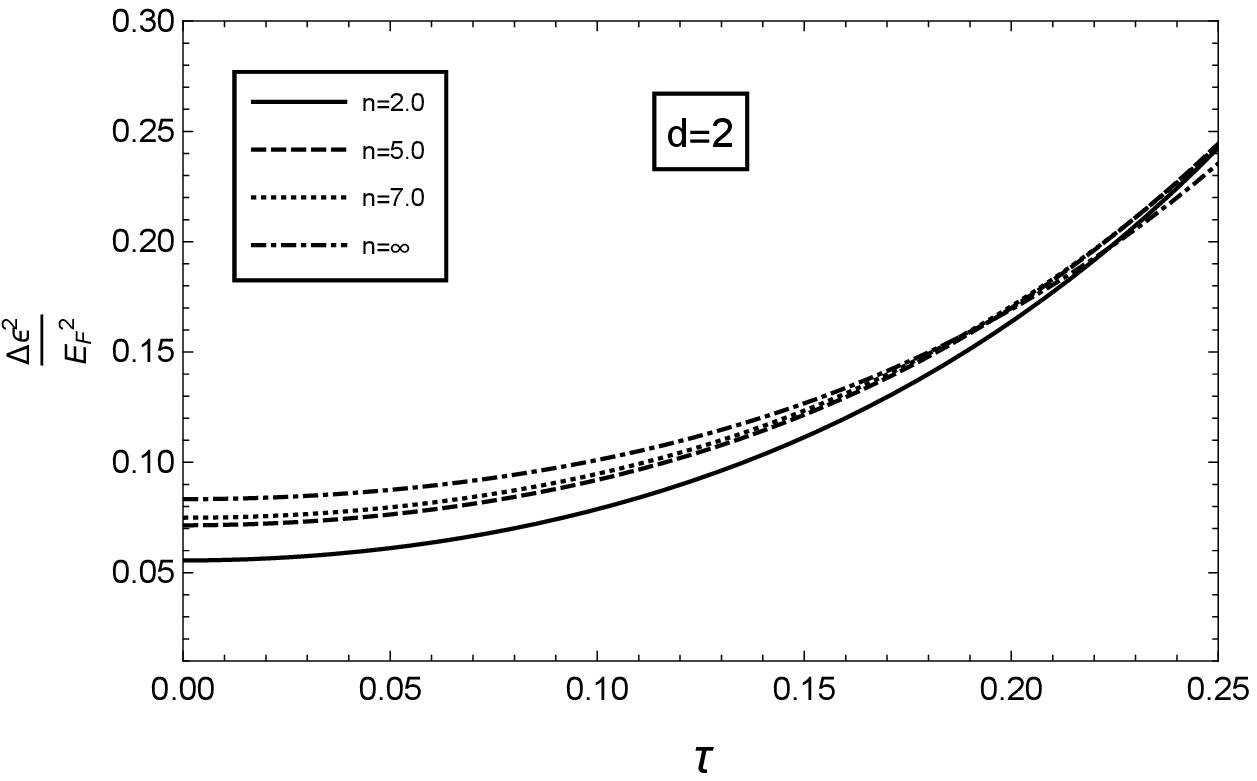}  
\includegraphics[ height=8cm, width=8cm]{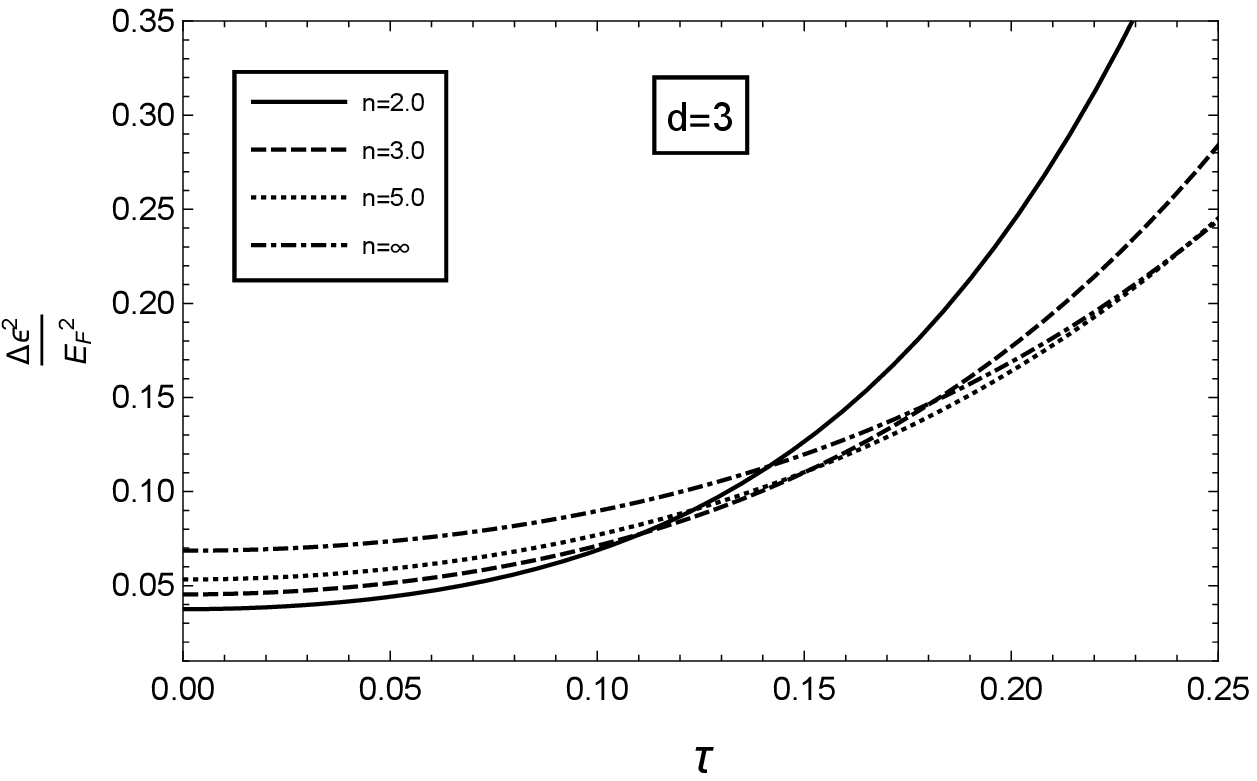}  
\caption{Energy fluctuation of ideal trapped Fermi gas
in $d=1$, $d=2$ and $d=3$. }
  \label{fig:boat1}
\end{figure}
 the energy fluctuation is zero at $T=0K$ for any value of $\chi$\cite{s,mms}. But,
interesingly, it is clear from the figure 1 that,
 the value of energy fluctuation at $T=0K$,  changes with varying $\chi$.
 We will explore this phenomena in detail, later. But, no hump is
 noted in the energy fluctuation of Fermi gas, unlike the Bose gas. This result 
 is in agreement with Biswas et. al.\cite{s}. \\\\
 It has already been reported,
 within the canonical ensemble that,
 energy fluctuation $\Delta \epsilon ^2$ is 
 related to specific heat $C_V$
 as $\Delta \epsilon ^2=kT^2 C_V$. And in the case of grand
 canonical ensemble this is true for ideal classical gas\cite{s} only.
Ref. \cite{s,mms}, shows  how the non zero fugacity of quantum gas
causes this relation to remain invalid for quantum system.
Now the quantum gases behave as a classical gas
in the high temperature limit and thus tend to maintain this relation at high temperature.
So, the status of this relation to be invalid is very 
important in low temprature limit. 
The low temperature limit of Bose gas corresponds to condensed phase. And the status of this relation
has already been checked for Bose system by Mehedi et. al. \cite{mms}. And it can be checked
for trapped Fermi gas in the degenerate limit
with the help of Eq. (9),
which depicts
the energy fluctuation  of trapped fermi gas changes as
$\Delta \epsilon ^2\sim A_1 + A_2 T^2-A_3 T^4$, while $C_V$ changes as $C_V\sim A T$\cite{mms}, 
where $A$, $A_i$ are functions of $\chi$. The temperature dependency of 
$\Delta \epsilon ^2$ and $C_V$ explicitly shows,
how $\Delta \epsilon ^2=kT^2 C_V$ relation is not maintained in low temperature limit of trapped Fermi gas.
 \\
\\
Let us further analyse the energy fluctuation in different space dimensions.
In  figure  2 we have set the condition of symmetric potential i.e. $n_1=n_2=....=n_d=n$. 
The non zero energy fluctuation at $T=0K$
is visible in all the figures.
But this non zero
value of energy fluctuation at $T=0K$, $\Delta \epsilon_0 ^2$
changes with different trapping potential. To be more specific, 
$\Delta \epsilon_0 ^2$
decreases with decreasing value  of 
$n$. When all $n_i=\infty$ (free system)
$\Delta \epsilon_0 ^2$
has the highest value and this value reduces as
we decrease $n$ (figure 2).
One can also find out from figure 2 (a)-(c) that,
$\Delta \epsilon_0 ^2$ also changes with dimensionality. In order to see this in detail we have done a 
separate plot.
Nevertheless, the influence of trapping potential is observed not only at $T=0K$ but also 
in the whole temperature range for any dimensionality.\\\\
\begin{figure}[h!]
\centering
\includegraphics[ height=10cm, width=10cm]{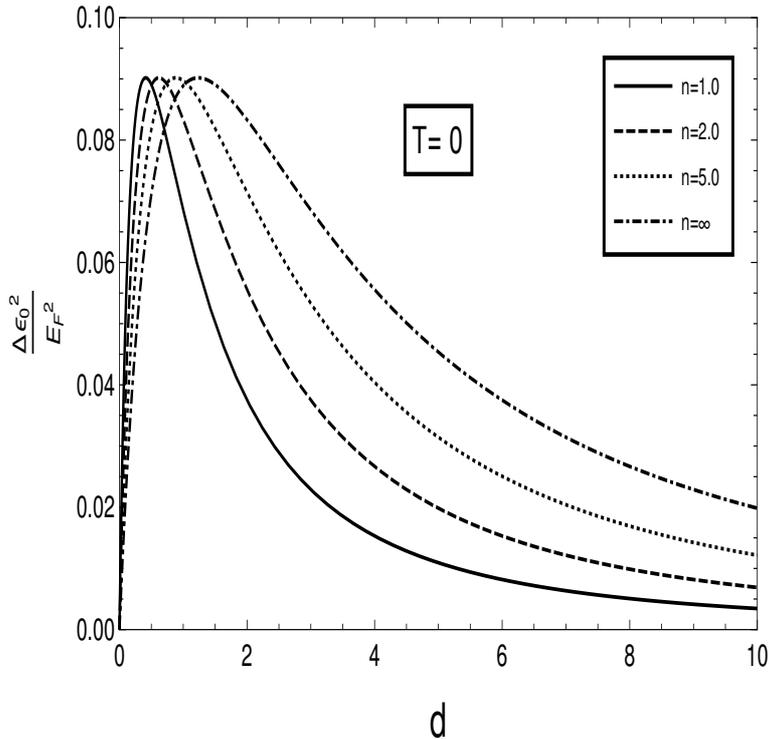}  
\caption{Change of energy fluctuation at $T=0K$ ($\Delta \epsilon_0 ^2$ )
with respect to dimensionality.}
  \label{fig:boat1}
\end{figure}
In figure 3, we analyse the change of 
$\Delta \epsilon_0 ^2$
with respect to dimensionality, 
for different types of trapping potential.
It is seen from the figure that,
for free Fermi system the 
$\Delta \epsilon_0 ^2$ is maximum near $d=1$. But this situation changes 
while the Fermi system is trapped,  as a shift is noticed in  the  maximum of
$\Delta \epsilon_0 ^2$.
Another important point to notice from figure 3 that,
the maximum of 
$\Delta \epsilon_0 ^2$
can be obtained in different dimensionality (depends on the trapping potential), 
but the maximum value of
$\Delta \epsilon_0 ^2$ 
remains the same for all. This
figure is very significant, as from it one can predict
which trapping potential will cause maximum 
of energy fluctuation at $T=0K$
 for any specific dimensionality.
 
\section{Conclusion}
In this manuscript we have restricted our discussion for
ideal nonrelativistic
fermi gases. Our general result on
energy fluctuation of ideal Fermi gas trapped under generic power law potential
in arbitrary dimension can reproduce exactly the same result in three dimension\cite{s}
but, it
 will be interesting
to calculate energy fluctuation
for
interacting fermions, which is not yet done.
Interaction might change the dependence of energy fluctuation on trapping
potential.
However, we are currently investigating the energy fluctuation for ideal relativistic
Fermi gases by taking into account the presence of antiparticles.
 \section{Acknowledgement}
 One of us (MMF) would like to thank Mr. Alamgir Al Faruqi for his cordial hospitality during MMF's stay 
 in London, UK.

 \end{document}